\documentclass[twocolumn]{svjour3} 

\usepackage[american]{babel}
\usepackage[utf8]{inputenc}
\usepackage{ae,aecompl}
\usepackage[numbers]{natbib}

\usepackage{bm}
\usepackage{color}
\usepackage{booktabs}
\usepackage{graphicx}
\usepackage{subfigure}

\usepackage{amsmath} 
\usepackage{amssymb}
\usepackage{commath}
\usepackage{dsfont}

\journalname{Meccanica}

\newcommand{\R}{\ensuremath{\mathbb{R}}}           

\newcommand{\indfunc}[1]{\mathds{1}_{#1}}

\DeclareMathOperator*{\argmax}{arg\,max}

\renewcommand{\SS}{\Theta}

\newcommand{\SSpt}{\theta}

\newcommand{\SF}{\Sigma}

\newcommand{\PM}{\mathcal{P}}

\newcommand{\pdf}[1]{p_{\tiny{#1}}}

\newcommand{\expval}[1]{E \left\lbrace   #1 \right\rbrace  }

\newcommand{\mean}[1]{\mu_{#1}}

\begin{document}

\title{Robust optimization and uncertainty quantification in the nonlinear mechanics of an elevator brake system}

\titlerunning{}

\author{Piotr Wolszczak \and Pawel Lonkwic \and Americo Cunha~Jr \and Grzegorz Litak \and Szymon Molski}
\authorrunning{P. Wolszczak \and P. Lonkwic \and A. {Cunha~Jr} \and G. Litak \and S. Molski}

\institute{P. Wolszczak \and G. Litak \at
				Lublin University of Technology,
				Faculty of Mechanical Engineering,
				Nadbystrzycka 36, PL-20-618 Lublin, Poland\\
				\email{p.wolszczak@pollub.pl}\\
				\email{g.litak@pollub.pl}
              \and
              P. Lonkwic \at
              State School of Higher Education,
		      Mechanical Engineering Faculty,
			  Pocztowa 54, PL-22-100 Chelm, Poland\\
              \email{plonkwic@gmail.com}
              \and
              A. Cunha~Jr \at
              Universidade do Estado do Rio de Janeiro,
		      Nucleus of Modeling and Experimentation with Computers -- NUMERICO,
			  Rua S\~{a}o Francisco Xavier, 524, Rio de Janeiro, 20550-900, RJ, Brazil\\
              \email{americo@ime.uerj.br}
              \and
               S. Molski \at
              AGH University of Science and Technology, Department of Rope Transport,
			  Mickiewicza 30, PL-30-059 Krakow, Poland\\
			  \email{molski@agh.edu.pl}
}

\date{Received: date / Accepted: date}

\maketitle

\begin{abstract}
This paper deals with nonlinear mechanics of an elevator brake system
subjected to uncertainties. A deterministic model that relates the braking force 
with uncertain parameters is deduced from mechanical equilibrium conditions.
In order to take into account parameters variabilities, a parametric probabilistic approach 
is employed. In this stochastic formalism, the uncertain parameters are modeled 
as random variables, with distributions specified by the maximum entropy principle. 
The uncertainties are propagated by the Monte Carlo method, which provides a 
detailed statistical characterization of the response. This work still considers the optimum design 
of the brake system, formulating and solving nonlinear optimization problems, with and without 
the uncertainties effects.


\end{abstract}

\section{Introduction}
\label{intro}

Considerations regarding the construction of lifting devices (design of cranes), 
and in particular brake systems, are not often discussed in the scientific literature,
{as the corresponding dynamical conditions are difficult to determine.}
The first person who addressed the issue of the impact of safety gears construction 
on the braking distance was Elisha Graves Otis, who in 1853 built the first safety {gears}
and subjected them to experimental studies \cite{LonkwicMono2017, Pater2011}. 
Subsequent works on this subject, {studying} several aspects of cranes mechanics,
appeared in the twentieth century {and were} published in journals and 
conference proceedings \cite{Yost1996p521,Lonkwic2015p363, Lonkwic2017p90,
Lonkwic2016p4401,Kaczmarczyk2006, Kaczmarczyk2009p012047,
 cunhajr2017proceng2}.

For instance, Yost and Rothenfluth \cite{Yost1996p521} describe how to configure
a lifting device and how to select the correct components. These issues constituted 
a significant contribution to the development of the configuration of lifts, ensuring a 
trouble-free operation.

Lonkwic \cite{Lonkwic2015p363} presents a comparative analysis of the operation 
of slip safety gear of his own design study with the models {by} leading European 
manufacturers. Deceleration (braking time) values obtained in the physical experiment 
are analyzed. In \cite{Lonkwic2017p90}, the same author and collaborators address, 
by means of wavelet analysis, how certain variables {influence} on the operating 
conditions of deceleration. A similar analysis is presented in \cite{Lonkwic2016p4401}, 
which {concerns the} selected braking parameters of CHP2000 and PP16 type chaters 
using the analysis of recursive patterns.

Regarding the study of elevator systems with uncertain operating conditions, the literature 
is not very comprehensive. The only works in this line known by the authors are
\cite{Kaczmarczyk2006,Kaczmarczyk2009p012047}, developed by Kaczmarczyk et al.,
{who} attempt to analyze the behavior of balance ropes due to harmonic and stochastic 
excitations, and Col\'{o}n et al. \cite{cunhajr2017proceng2}, {who calculate} the propagation 
of the rail profile uncertainties and study the effectiveness of a closed-loop control law.

Even with the scientific literature being rich in studies regarding the behavior of vehicles
brake systems under changing operating conditions \cite{Renault2016p37,Dezi2014p1011,Knops2006p693}, 
it is surprising that, to the best of authors' knowledge, no similar research description on lift brakes 
{has} been reported up to the present date. {Only} general provisions contained 
in the British Standard Document BS EN 81 \cite{EN8120, EN8150} {are to be found}.

Thus, seeking to fill this gap, the present work aims to study the influence of some
operating conditions on {the efficiency of an elevator brake device, by analyzing how the
operating parameters underlying uncertainties propagate through the mechanical system.}
In particular, the cam brake angle and the spring reaction force {are of interest}.
In addition to {quantifying} the effects of uncertainties in operating conditions, this study 
also aims to achieve a {robust} design of a brake system by solving 
a nonlinear optimization problem, considering (or not) {the uncertainty} effects.

The {remaining part} of this paper is organized as follows: the deterministic modeling of the
lift brake systems under study is presented in section 2. In section 3, 
the construction of a consistent stochastic model of uncertainties to deal with 
variabilities in the uncertain parameters{, is presented}. Two optimization problems, one classical
and one robust, which seek to find an optimal {design} for the brake system
are {formulated} in section 4. In section 5, numerical experiments are reported and 
discussed. Finally, in section 6, concluding remarks are presented.


\section{Deterministic modeling}
\label{determ_model}

\subsection{Elevator brake system}

\begin{figure}[h!]
\centering
\includegraphics[scale=0.4]{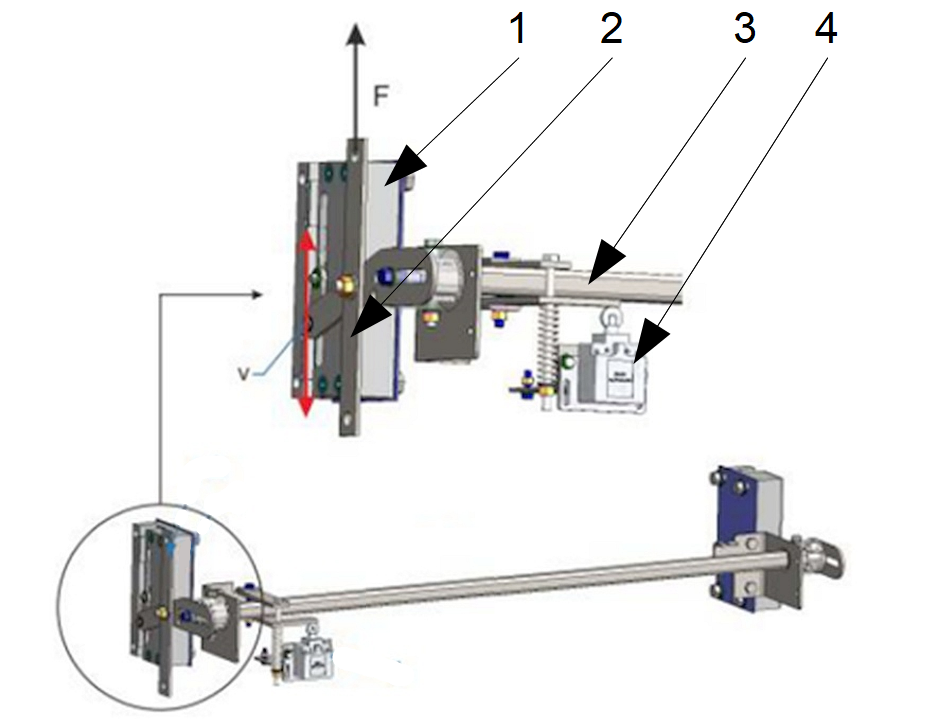}
\caption{Schematic representation of the CHP 2000 safety gear used by a friction crane brake system,
which consists of the following parts: 1 - steel body; 2 - brake roller; 3 - thrust plate; 4 - spring package; 
5 - braking cam; 6 - lift guide.}
\label{CHP2000_fig}
\end{figure}

{
A schematic representation of the CHP 2000 safety gear, used by a typical friction crane brake system, is presented 
in Figure~\ref{CHP2000_fig}. 
It consists of a monolithic steel body \textcolor{blue}{(1)}, in which a braking cam \textcolor{blue}{(5)} is mounted on a bolt. 
The braking cam moves the brake roller \textcolor{blue}{(2)}, which has a knurled surface. 
This irregular surface is responsible for the braking process and for the cooperation with the 
guide roller surface. The brake roller moves over the braking cam surface until it contacts the lift guide \textcolor{blue}{(6)}. 
It is a free movement that does not cause any braking effect. The second part of the braking 
process is in constant contact with the lift guide surface.
}

\begin{figure}[h!]
\centering
\includegraphics[scale=0.3]{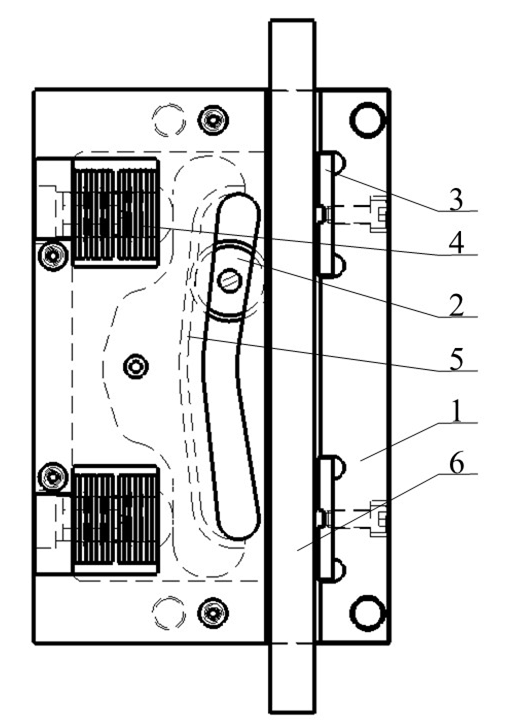}
\caption{Illustration of a typical friction crane brake system used by lifting devices,
which consists of the following parts: 1 - safety gear cabin; 2 - trigger lever; 
3 - safety gears connector; 4 - rope speed limiter.}
\label{crane_break_fig}
\end{figure}

{
An illustration of a typical friction crane brake system used by lifting devices 
is shown in Figure~\ref{crane_break_fig}, which indicates the different components 
of the mechanism (see the caption). It consists of two safety gears, moving on the lift guides, 
connected to each other to ensure simultaneous operation when the brakes are activated 
by means of a trigger lever. A lift safety gear is placed in the frame, under a safety gear cabin. Its trigger
is attached to the trigger lever, in which the end is connected to the rope speed limiter.
In the upper part of the elevator shaft there is a speed limiter supervising the work of the safety gear, 
and in its lower part load responsible for causing the proper tension of the speed 
limiter rope {is located}.  The speed limiter triggers the braking process when the nominal speed of the elevator car is 
increased by 0.3 m/s. After exceeding the nominal speed, the speed limiter is blocked, and the 
rope is also immobilized.
}

{
During the movement of an elevator car with locked components, the lever is moved in the opposite 
direction {to} the cabin, triggering the brake safety gear roll. In its turn, the roll is pressed against the 
guide causing elastic deformation towards the thrust plate located on the other side of the disk spring
package, which induces the loss of energy in the accelerating mass. Therefore, the disc spring package is 
responsible for a variable force that presses the roller to the guide during the braking process.
}

\subsection{Mathematical model}

The design assumptions and safety gear structure shown 
in Figure~\ref{CHP2000_fig} are {taken} into account
to construct a mathematical model that relates 
the braking force with geometric parameters and other 
characteristics of the mechanical system. In this sense,
equilibrium conditions for the system are deduced below.

\begin{figure}[h!]
\centering
\includegraphics[scale=1.5]{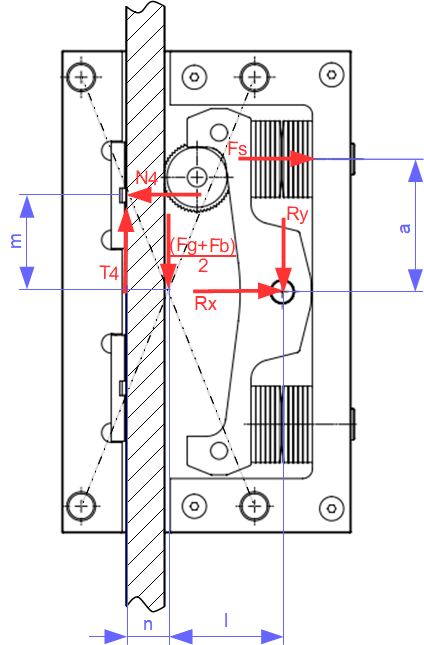}
\caption{Illustration of the forces (in red) acting on the safety 
gear steel body and the underlying geometric dimensions (in blue).}
\label{trunk_fig}
\end{figure}

A free-body diagram can be seen in 
Figure~\ref{trunk_fig}, which shows a schematic representation 
of the forces (in red) acting on the safety gear steel body, and 
the underlying geometric dimensions (in blue).

A balance of forces and {the} moments acting on the steel 
body gives rise to the equations
\begin{equation}
-N_{4} + F_{s} + R_{x} = 0,
\label{trunk_equil_x}
\end{equation}
\begin{equation}
T_{4} - R_{y} - (F_g+F_b)/2=0,
\label{trunk_equil_y}
\end{equation}
\begin{equation}
- F_{s} \, a - R_{y} \, l + N_{4} \,  m  - T_{4} \, n = 0,
\label{trunk_equil_mom}
\end{equation}
where $F_{s}$ is the spring reaction force;
$F_{b}$ is the inertial force from the cabin and lifting capacity;
$F_{g}$ is the cabin and lifting capacity weight;
$T_{4}$ is the friction force between the guide and brake retaining block, and
$N_{4}$ is the corresponding normal force;
$R_{x}$ and $R_{y}$ are the reaction forces in the braking cam rotation point;
while $a$, $l$, $m$ and $n$ are geometric
dimensions {depicted} in Figure~\ref{trunk_fig}.

\begin{figure}[h!]
\centering
\includegraphics[scale=1.5]{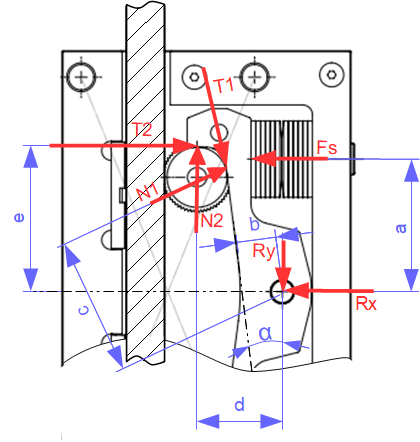}
\caption{Illustration of the forces (in red) acting on the wedge
and the underlying geometric dimensions (in blue)
and }
\label{wedge_fig}
\end{figure}

The forces (in red) acting on the wedge during braking and immediately after stopping 
the cabin, until the safety gears are unlocked by technical maintenance of the lift, 
are shown in Figure~\ref{wedge_fig}, {along with} the relevant geometric dimensions (in blue).

A new balance of forces and moments provides
\begin{equation}
T_{2} + N_{1} \, \cos{\alpha} + T_{1} \, \sin{\alpha} - F_{s} - R_{x} = 0,
\label{wedge_equil_x}
\end{equation}
\begin{equation}
F_{s} \, a + T_{1} \, b - N_{1} \, c - N_{2} \, d - T_{2} \, e =0,
\label{wedge_equil_mom}
\end{equation}
where
$T_{1}$ and $T_{2}$ are friction forces between brake elements (roller and cam),
$N_{1}$ and $N_{2}$ are the corresponding normal forces;
$\alpha$ is the braking cam angle; $b$, $c$, $d$ and $e$ are other geometric
dimensions of the problem, shown in Figure~\ref{wedge_fig}.

\begin{figure}[h!]
\centering
\includegraphics[scale=1.5]{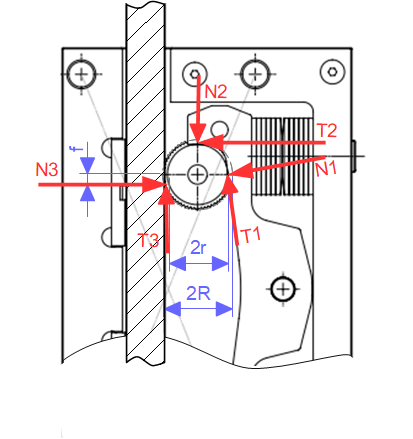}
\caption{Illustration of the forces (in red) acting on the brake roller 
inside the safety gear and and the underlying geometric dimensions (in blue).}
\label{roll_fig}
\end{figure}

In Figure~\ref{roll_fig} the reader can see characteristic dimensions (in blue)
and forces (in red) acting on the brake roller inside the safety gear. 

Now the balance of forces gives
\begin{equation}
N_{3}  - T_{2} - N_{1} \, \cos{\alpha} - T_{1} \,  \sin{\alpha} = 0,
\label{roll_equil_x}
\end{equation}
\begin{equation}
T_{3}  - N_{2} + T_{1} \, \cos{\alpha} - N_{1} \,  \sin{\alpha} = 0,
\label{roll_equil_y}
\end{equation}
where $T_{3}$ and $N_{3}$ respectively denotes the frictional and the normal
forces between brake roller and the guide.

The frictional forces $T_1$, $T_2$ and $T_4$ are, respectively, 
related to the normals $N_1$, $N_2$ and $N_4$ through 
a Coulomb friction model, so that
\begin{equation}
T_{1}  = \mu_{1} \, N_{1},
\label{friction_eq1}
\end{equation}
\begin{equation}
T_{2}  = \mu_{2} \, N_{2},
\label{friction_eq2}
\end{equation}
\begin{equation}
T_{4}  = \mu_{4} \, N_{4},
\label{friction_eq4}
\end{equation}
where $\mu_1$, $\mu_2$ and $\mu_4$ are friction coefficients.

On the other hand, the relationship between the frictional force $T_3$ 
and the normal $N_3$ takes into account the plastic deformation occurring 
in the contact between brake roller and the guide, so that
\begin{equation}
T_{3}  = \frac{f}{R} \, N_{3},
\label{friction_eq3}
\end{equation}
where $f$ and $R$ are geometric dimensions defined in
Figure~\ref{roll_fig}.

The vertical reaction force $R_{y}$ can be obtained from
Eqs.(\ref{trunk_equil_y}) and (\ref{trunk_equil_mom}), 
\begin{equation}
R_{y} = - T_{4} + \frac{(F_g+F_b)}{2},
\end{equation}
\begin{equation}
R_{y} = \frac{F_{s} \, a - N_{4} \,  m  + T_{4} \, n}{l},
\end{equation}
which, when combined together with Eq.(\ref{friction_eq4}),
allows one to express $N_4$ as
\begin{equation}
N_4 = \frac{ (F_g+F_b) \, l/2 - F_s \, a}{\mu_4 \, (n+l) - m}.
\label{N4_eq}
\end{equation}

Similarly, from suitable manipulations of Eqs.(\ref{roll_equil_x}), (\ref{friction_eq1}) 
and (\ref{friction_eq2}), it can be concluded that
\begin{equation}
N_3 = \mu_2 \, N_2 + (\mu_1 \, \sin{\alpha} + \cos{\alpha}) \, N_1,
\label{N3_eq}
\end{equation}
as well as, from Eqs.(\ref{wedge_equil_mom}), (\ref{friction_eq1})
and (\ref{friction_eq2}), it is possible to obtain
\begin{equation}
N_2 = \frac{a \, F_s + (b \, \mu_1 -c ) \, N_1}{d + e \, \mu_2},
\label{N2_eq}
\end{equation}
which, in combination with Eqs.(\ref{trunk_equil_x}),
(\ref{wedge_equil_x}), (\ref{friction_eq1}) and (\ref{friction_eq2}),
gives rise to
\begin{equation}
N_1 = \frac{N_4 - \displaystyle \frac{a \, \mu_2 \, F_s}{d + e \, \mu_2}}{\mu_1 \, \sin{\alpha} + \cos{\alpha} + \displaystyle \frac{\mu_2 \, (b \, \mu_1 - c)}{d + e \, \mu_2}}.
\label{N1_eq}
\end{equation}

The braking force, {resulting} from the joint superposition of all 
frictional forces, is given by
\begin{equation}
F_h = T_1 + T_2 + T_3 + T_4,
\label{Fh_eq1}
\end{equation}
which, with aid of Eqs.(\ref{friction_eq1}) to (\ref{friction_eq3}),
can be {rewritten} as
\begin{equation}
F_h = \mu_1 \, N_1 + \mu_2 \, N_2 + \frac{f}{R} \, N_3 + \mu_4 \, N_4.
\label{Fh_eq2}
\end{equation}

Note that, once normal forces $N_1$, $N_2$, $N_3$ and $N_4$
present {explicit} dependence on geometric dimensions, frictional {coefficients},
and non-frictional forces, the braking force $F_h$ is also a function of these
parameters, i.e., 
\begin{multline}
F_h = F_h(\alpha, F_s, F_g, F_b, \mu_1, \mu_2, \mu_4, \cdots \\
\cdots a, b, c, d, e, f, l, m, n, R).
\label{Fh_eq3}
\end{multline}


\section{Stochastic modeling}
\label{stoch_model}

The angle $\alpha$ and the spring reaction
force $F_s$ are {subjected} to variabilities during the
operation conditions of the brake system, so that
their actual values may be very different from the 
nominal project values.
{Since} they are {the} critical parameters for the brake system efficiency, 
{studying} the effect of such variabilities on the braking force is essential for a good design. 
In this way, a parametric probabilistic approach 
\cite{cunhajr2017,soize2017} is employed here
to construct a consistent stochastic model
for uncertain parameters $\alpha$ and $F_s$.

\subsection{Probabilistic framework}

Let $(\SS, \SF, \PM)$ be the probability space used to
describe the model parameters uncertainties {\cite{soize2017,soize2013p2379}},
where $\SS$ is the sample space, $\SF$ a $\sigma$-field over $\SS$, 
and $\PM: \SF \to [0,1] $ a probability measure.

In this probabilistic setting, the parameters $\alpha$ and $F_s$
are respectively described by the random variables $X_1: \SS \to \R$
and $X_2: \SS \to \R$, which are lumped into the random vector
$\bm{X}: \SS \to \R^2$, which associates to each elementary event
$\SSpt \in \SS$ a vector $\bm{X} = (X_1, X_2)$. 
The probability distribution of $\bm{X}$ is characterized by 
the map $\pdf{\bm{X}}: \R^2 \to [0, + \infty)$, dubbed the joint 
probability density function (PDF).

The mean value of $\bm{X}$ is defined in terms of the expected value operator
\begin{equation}
	\expval{\bm{X}} = \int \, \int_{\R^2} \bm{x} \, \pdf{\bm{X}}(\bm{x}) \, d \bm{x} \, ,
	\label{def_expval_op}
\end{equation}
in which $\bm{x}=(x_1, x_2)$ and $d \bm{x} = dx_1 \, dx_2$.

\subsection{Maximum entropy principle}

To perform a judicious process of uncertainty quantification, 
it is essential to construct a consistent stochastic 
model for the random vector $\bm{X}$, that represents
the uncertainties in $\alpha$ and $F_s$ in a rational way, 
trying to be {unbiased} as possible.
In this sense, in order to avoid possible physical inconsistencies 
in the probabilistic model, only available information must be used 
in its construction \cite{soize2017,soize2013p2379}.
{
When this information materializes in the form of a large set of 
experimental data, the standard procedure is to use a nonparametric 
statistical estimator to infer the joint distribution of $\bm{X}$ \cite{soize2017,soize2013p2379}.
However, if} little (or even {no}) experimental data for $X_1$ and $X_2$ is available, 
as is the case {of} this paper, such construction can be done 
based only on known theoretical information, with {the} aid of the 
\emph{maximum entropy principle} \cite{soize2017,soize2013p2379}.

The available theoretical information about the random parameters
$X_1$ and $X_2$ encompasses a range of possible values
for each of then, i.e.,
\begin{equation}
	(X_1, X_2) \in [\alpha_1, \alpha_2] \times [\gamma_1, \gamma_2]
	\subset [0, 2\pi] \times (0, + \infty),
\end{equation}
as well as their nominal values $\mean{X_1}$ and $\mean{X_2}$, 
that are assumed {to be} equal to their mean values, i.e.,
\begin{equation}
	\expval{X_1} = \mean{X_1} \in [\alpha_1, \alpha_2],
\end{equation}
\begin{equation}
	\expval{X_2} = \mean{X_2} \in [\gamma_1, \gamma_2].
\end{equation}

This information is translated into the statistical language through
the normalization condition
\begin{equation}
	\int_{\gamma_1}^{\gamma_2} \int_{\alpha_1}^{\alpha_2} \pdf{\bm{X}} (\bm{x}) \, d \bm{x} \, = 1,
	\label{MaxEnt_eq1}
\end{equation}
and the first order moment equation
\begin{equation}
	\int_{\gamma_1}^{\gamma_2} \int_{\alpha_1}^{\alpha_2} \bm{x} \, \pdf{\bm{X}} (\bm{x}) \, d \bm{x} \,
	= \left( \mean{X_1}, \mean{X_2} \right).
	\label{MaxEnt_eq2}
\end{equation}

From the information theory point of view, the most rational approach 
to specify the distribution of $\bm{X}$ in this scenario of reduced information 
is through the maximum entropy principle (MaxEnt) \cite{soize2017,soize2013p2379,kapur1992}, 
which seeks the PDF that maximizes the entropy functional
\begin{equation}
	S(\pdf{\bm{X}}) = - \int_{\gamma_1}^{\gamma_2} \int_{\alpha_1}^{\alpha_2} 
	\pdf{\bm{X}} (\bm{x}) \, \ln{\pdf{\bm{X}} (\bm{x})} \, d \bm{x} \, ,
	\label{MaxEnt_entropy}
\end{equation}
respecting the restrictions (information) defined by (\ref{MaxEnt_eq1}) and (\ref{MaxEnt_eq2}).

Using the Lagrange multipliers method it is possible to show that such joint PDF is given by
\begin{equation}
	\pdf{\bm{X}} (\bm{x}) = \pdf{X_1} (x_1) \times \pdf{X_2} (x_2),
	\label{MaxEnt_pdf}
\end{equation}
with marginal densities
\begin{equation}
	\pdf{X_1} (x_1) = \exp{\left( - \lambda_{10} - \lambda_{11} \, x_1 \right)} \, \indfunc{[\alpha_1, \alpha_2]} (x_1),
	\label{MaxEnt_pdf1}
\end{equation}
\begin{equation}
	\pdf{X_2} (x_2) = \exp{\left( - \lambda_{20} - \lambda_{21} \, x_2 \right)} \, \indfunc{[\gamma_1, \gamma_2]} (x_2),
	\label{MaxEnt_pdf2}
\end{equation}
where {$\lambda_{10}$, $\lambda_{11}$, $\lambda_{20}$ and $\lambda_{21}$}
are parameters of the distribution of $\bm{X}$, and
\begin{equation}
	\indfunc{I}(x) = 
	\begin{cases}
       1, & x \in I,\\
       0, & x \not \in I,\\ 
     \end{cases}
\end{equation}
denotes the indicator function of the interval $I$. Note that, since no information 
relative to the cross statistical moments between $X_1$ and $X_2$ has been 
provided, MaxEnt provides independent distributions.

The parameters {$\lambda_{10}$, $\lambda_{11}$, $\lambda_{20}$ and $\lambda_{21}$} depend on 
$\alpha_1$, $\alpha_2$, $\gamma_1$, $\gamma_2$, $\mean{X_1}$ and 
$\mean{X_2}$. They are computed through {the nonlinear system 
of equations obtained by replacing 
(\ref{MaxEnt_pdf}) in (\ref{MaxEnt_eq2}) and in the normalization conditions of
the marginal PDFs (\ref{MaxEnt_pdf1}) and (\ref{MaxEnt_pdf2}).}

In a scenario with little information, it is practically impossible not to be biased 
in choosing a probability distribution. The MaxEnt formalism provides the least 
biased distribution that is consistent with the known information, {therefore constituting} the most 
rational approach \cite{kapur1992}.

\subsection{Uncertainty propagation}

The mathematical model {relating} the braking force $F_h$
with braking cam angle $\alpha$ and the spring reaction force $F_s$,
Eq.(\ref{Fh_eq3}), can be thought abstractly as a nonlinear deterministic 
functional $\mathcal{M}$ that maps a vector of input parameters $\bm{x}=(\alpha, F_s)$ 
into a scalar quantity of interest $y=F_h$, i.e., 
\begin{equation}
	\bm{x} \mapsto y = \mathcal{M}(\bm{x}).
	\label{model_eq}
\end{equation}

Thus, if the uncertain parameters $\alpha$ and $F_s$ are represented by 
the known random vector $\bm{X}$, the braking force becomes the random 
variable $Y= \mathcal{M}(\bm{X})$, for which the distribution must be estimated.

The process of determining the distribution of $Y$, once the probabilistic 
law of $\bm{X}$ is known, is dubbed uncertainty propagation problem 
\cite{cunhajr2017,soize2017}, 
being addressed in this paper via {the} Monte Carlo simulation
\cite{kroese2011,cunhajr2014p1355}.

In this stochastic calculation technique, $\nu$ independent samples of $\bm{X}$ 
are drawn according to the density (\ref{MaxEnt_pdf}),
giving rise to statistical realizations
\begin{equation}
	\bm{X}^{(1)}, \bm{X}^{(2)}, \cdots, \bm{X}^{(\nu)}.
\end{equation}
Each of these scenarios for $\bm{X}$ is given as input to the
nonlinear deterministic map $\bm{x} \mapsto y = \mathcal{M}(\bm{x})$, resulting 
in a set of possible realizations for the quantity of interest
\begin{equation}
	Y^{(1)}, Y^{(2)}, \cdots, Y^{(\nu)},
\end{equation}
where $Y^{(j)} = \mathcal{M}(\bm{X}^{(j)}), ~j=1, \cdots, \nu$.
These samples are used to estimate statistics of $Y$
{non-parametrically}, i.e., without assuming the PDF shape known 
\cite{wasserman2007}.


\section{Optimization framework}

Regarding the improvement of brake system efficiency, 
an optimal design of its components is required. 
This work addresses this question by solving nonlinear 
optimization problems that seeks to maximize the braking force, 
using geometric dimensions of the system as design variables.

Two optimization approaches are employed.
The first one, named \emph{classical}, is based {on} deterministic 
formalism of nonlinear programming \cite{Bonnans2009}, 
while the latter, dubbed \emph{robust}, takes into account the model 
parameters uncertainties, in order to reduce the optimum point sensitivity 
to small disturbances \cite{beyer2007p3190,Gorissen2015p124}.

In this framework, a set of two design variables (geometric dimensions) 
{is} denoted generically by the vector $\bm{s}$. The other parameters of the 
model are denoted generically by $\bm{x}$, and the model response
is given by the nonlinear map $(\bm{s}, \bm{x}) \mapsto y = \mathcal{M}(\bm{s}, \bm{x})$
The quantity of interest to be optimized (objective function) is denoted generically 
by $\mathcal{J}$.

\subsection{Classical optimization}

In this classical optimization approach {the} $\bm{s}$ {components} are 
employed as {design} variables, while the braking force is adopted 
as objective function, i.e.,
\begin{equation}
\mathcal{J}_{C} (\bm{s}) = y.
\label{opt_eq1}
\end{equation}

The admissible set for this optimization problem is defined by
$\mathcal{A}_{C} = [s_1^{min}, s_1^{max}] \times [s_2^{min}, s_2^{max}]$,
so that it can be formally stated as find an optimal design vector
\begin{equation}
\bm{s}_{C}^{opt} = \argmax_{\bm{s} \in \mathcal{A}_{C}} \, \mathcal{J}_{C} (\bm{s}).
\label{opt_problem1}
\end{equation}

\subsection{Robust optimization}

In this robust optimization framework, which is based 
on those shown in \cite{cunhajr2015p849,cunhajr2018bsme},
the uncertainties are described according to the formalism of the
section~\ref{stoch_model}, where $\bm{x}$ becomes the
random vector $\bm{X}$, and, as a consequence, 
$y = \mathcal{M}(\bm{s}, \bm{x})$ becomes the random variable 
$Y = \mathcal{M}(\bm{s}, \bm{X})$.

Thus, the robust objective function is not constructed directly 
from the model response, but with {the} aid of statistical measures
of $Y$, which aims to guarantee greater stability to small disturbances
(robustness) to an optimum point. 

Specifically, the robust objective function is given
by a convex combination between minimum, 
maximum, mean and standard deviation inverse,
so that
\begin{multline}
	\mathcal{J}_{R} (\bm{s}) = \beta_1 \, \min_{\SSpt \in \SS} \left\lbrace Y \right\rbrace +
										   \beta_2 \,  \max_{\SSpt \in \SS} \left\lbrace Y \right\rbrace + \\
											\beta_3 \, \expval{Y} +
										    \beta_4 \, \frac{1}{\sqrt{\expval{Y^2} -\expval{Y}^2}},
	\label{ropt_eq1}
\end{multline}
where $\beta_1+\beta_2+\beta_3+\beta_4=1$. 

Note that by maximizing this 
robust objective function, it is sought to raise both the lowest and the highest 
possible value, the mean, in addition to reducing the dispersion, by reducing 
the standard deviation.

Additionally, in order to avoid excessively small braking forces, 
the following probabilistic constraint is imposed
\begin{equation}
	\PM \left\lbrace |Y| > y^{\ast} \right\rbrace \geq 1 - P_r,
	\label{ropt_eq2}
\end{equation}
where $y^{\ast}$ is a lower bound for the magnitude of $Y$, and
$P_r$ is reference probability.

Therefore, the admissible set for the robust optimization problem,
denoted by $\mathcal{A}_{R}$, is defined as the subset of 
$\mathcal{A}_{C}$ for which the probabilistic constraint (\ref{ropt_eq2})
is respected. In this way, the robust optimization problem is formally 
defined as find an optimal design vector
\begin{equation}
\bm{s}_{R}^{opt} = \argmax_{\bm{s} \in \mathcal{A}_{R}} \, \mathcal{J}_{R} (\bm{s}).
\label{opt_problem2}
\end{equation}


\section{Results and discussion}
\label{num_exp}

The simulations reported below, {conducted in Matlab},
use the following numerical values for the deterministic 
parameters of the mechanical model:
$F_g = 50$ kN; $F_b = 30$ kN; 
$\mu_1 = 0.10$; $\mu_2 = 0.10$;
$\mu_4 = 0.15$;
$a = 55.0$ mm; $b = 16.6$ mm;
$c = 52.7$ mm; $d = 34.5$ mm;
$e = 60.7$ mm; $f = 0.005$ mm;
$l = 49.0$ mm; $m = 40.0$ mm;
$n = 17.5$ mm; $R = 29.0$ mm.

Regarding the two random parameters, 
the following information is assumed:
$[\alpha_1, \alpha_2] = [0,18] º$;
$[\gamma_1, \gamma_2] = [0, 56]$ kN;
$\mean{X_1} = 6º$; and $\mean{X_2} = 42$ kN.

\subsection{Uncertainty quatification}

\begin{figure}
\centering
\includegraphics[scale=0.5]{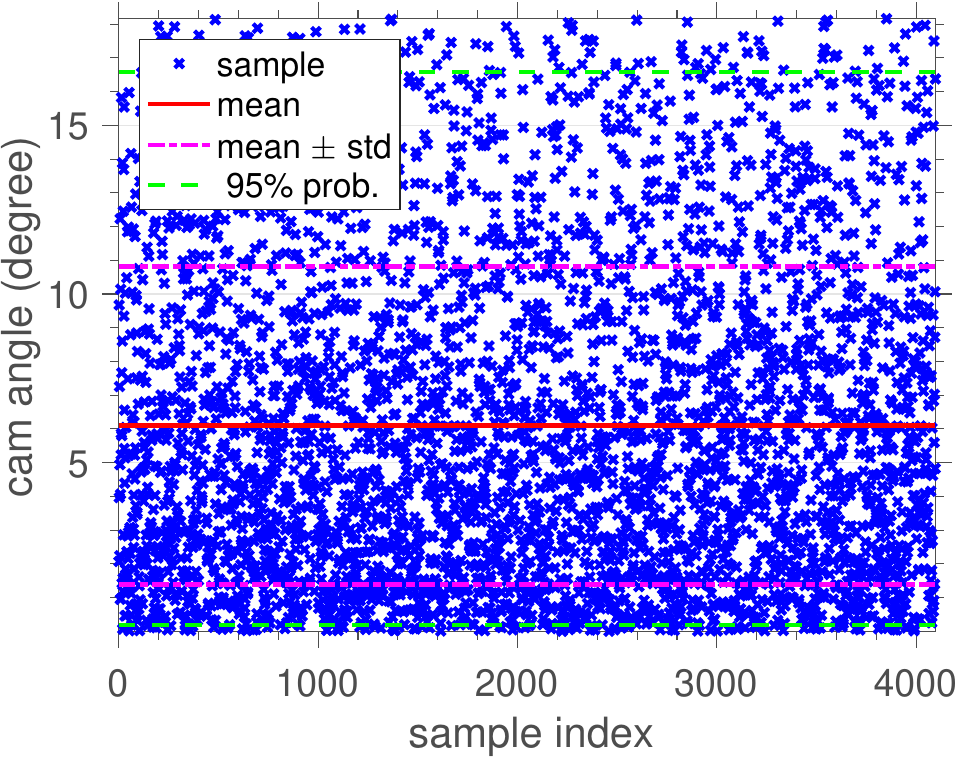} ~ 
\includegraphics[scale=0.5]{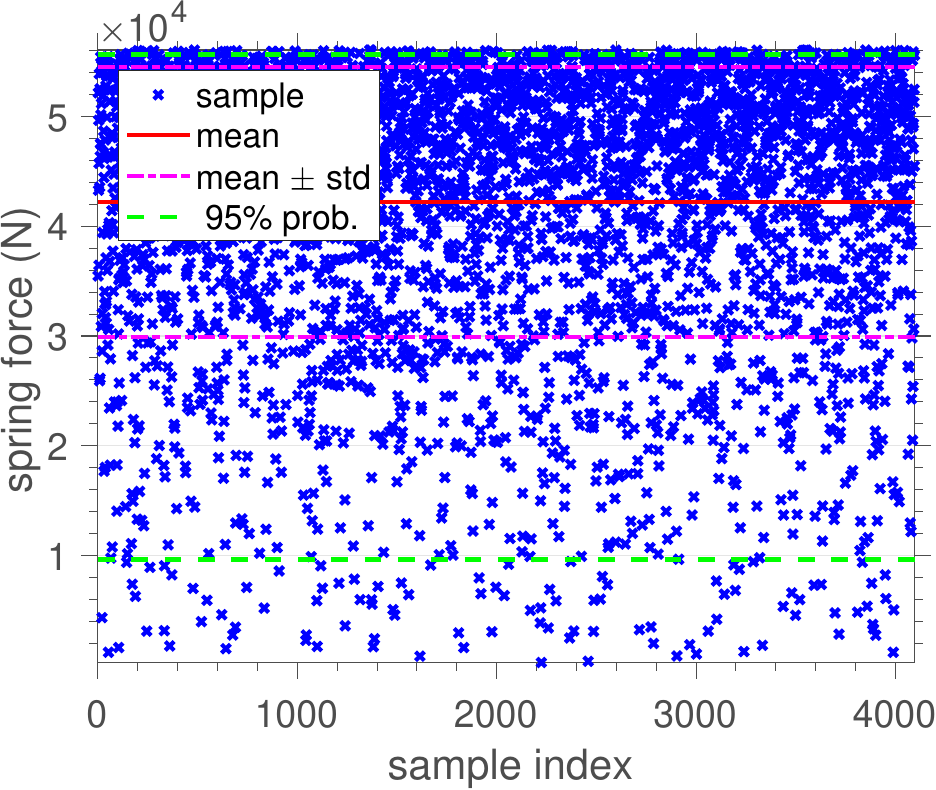}
\caption{Generated samples and statistical measures for 
the cam angle (top) and the spring force reaction (bottom).}
\label{input_samples_fig}
\end{figure}

The calculation of the propagation of uncertainties 
of $\bm{X}=(X_1, X_2)$ through the mechanical-mathematical 
model (\ref{model_eq}) initially involves the generation 
of random samples according to the probabilistic model 
defined by Eq.(\ref{MaxEnt_pdf}). A set of 4096 random 
samples for $X_1$ (top) and $X_2$ (bottom) can be seen 
in Figure~\ref{input_samples_fig}, which also shows some 
statistics (mean, standard deviation and 95\% confidence interval)
for this set of values.

\begin{figure}
\centering
\includegraphics[scale=0.5]{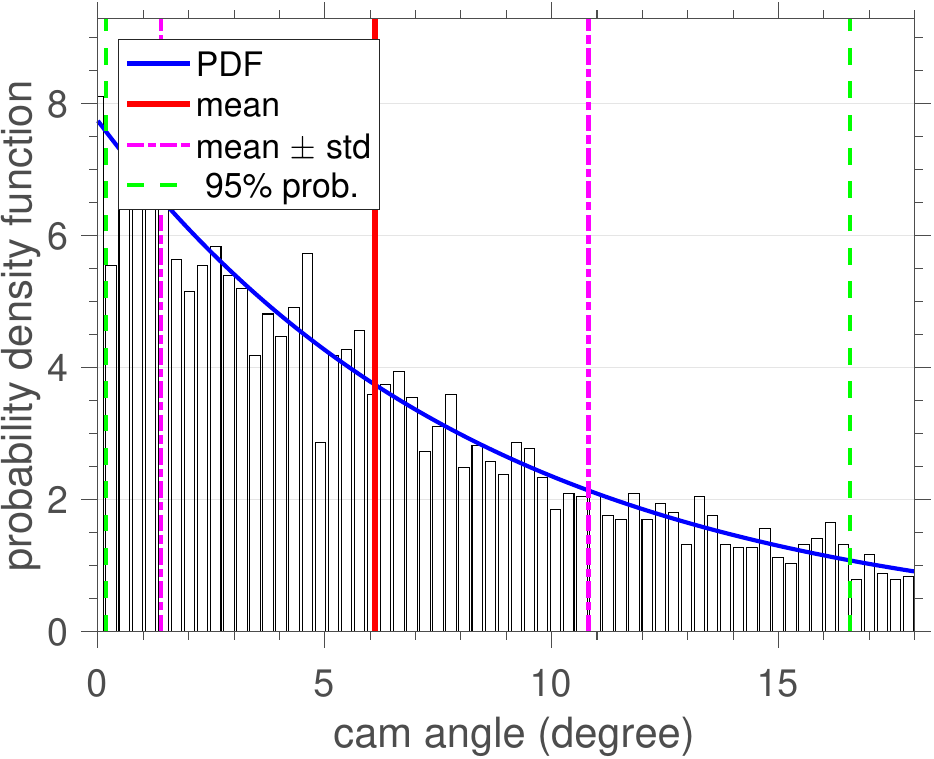} ~ 
\includegraphics[scale=0.5]{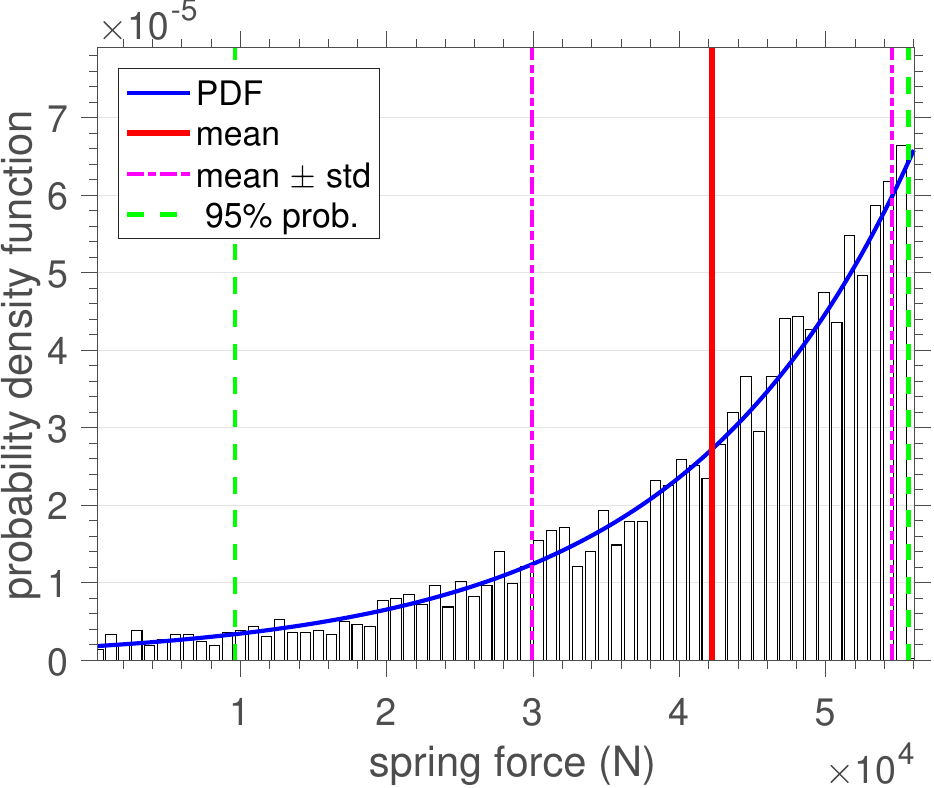}
\caption{Probability density functions and statistical measures for the cam angle (top) 
and the spring force reaction (bottom).}
\label{input_pdfs_fig}
\end{figure}

In Figure~\ref{input_pdfs_fig} the reader can see the
statistics shown in Figure~\ref{input_samples_fig} compared 
to the analytical curves for the PDFs of $X_1$ and $X_2$, 
and histograms constructed with the underlying random samples. 
It {can be observed} that the sampling process is well conducted, 
since the histograms and analytical curves present great similarity.

Note that the brake cam angle $X_1$ is modeled according to 
a probability density with a descending exponential behavior, 
which decays slowly between the ends of the support 
$[\alpha_1, \alpha_2] = [0, 18] º$, 
whereas the spring reaction force $X_2$ is described by probabilistic law 
with an increasing exponential density, which grows rapidly from 
the left to the right extreme of $[\gamma_1, \gamma_2] = [0, 56]$ kN.

It is worth noting that, of course, the real system parameters do not follow 
these probability distributions. These are only approximations of the real 
distributions, constructed with {the} aid of the maximum entropy principle and
the available information about these parameters. However, as in this 
paper the authors do not have experimental data to infer the real form of 
these distributions, in the light of information theory, the PDFs of 
Figure~\ref{input_pdfs_fig} are the best that can be inferred.

\begin{figure}
\centering
\includegraphics[scale=0.5]{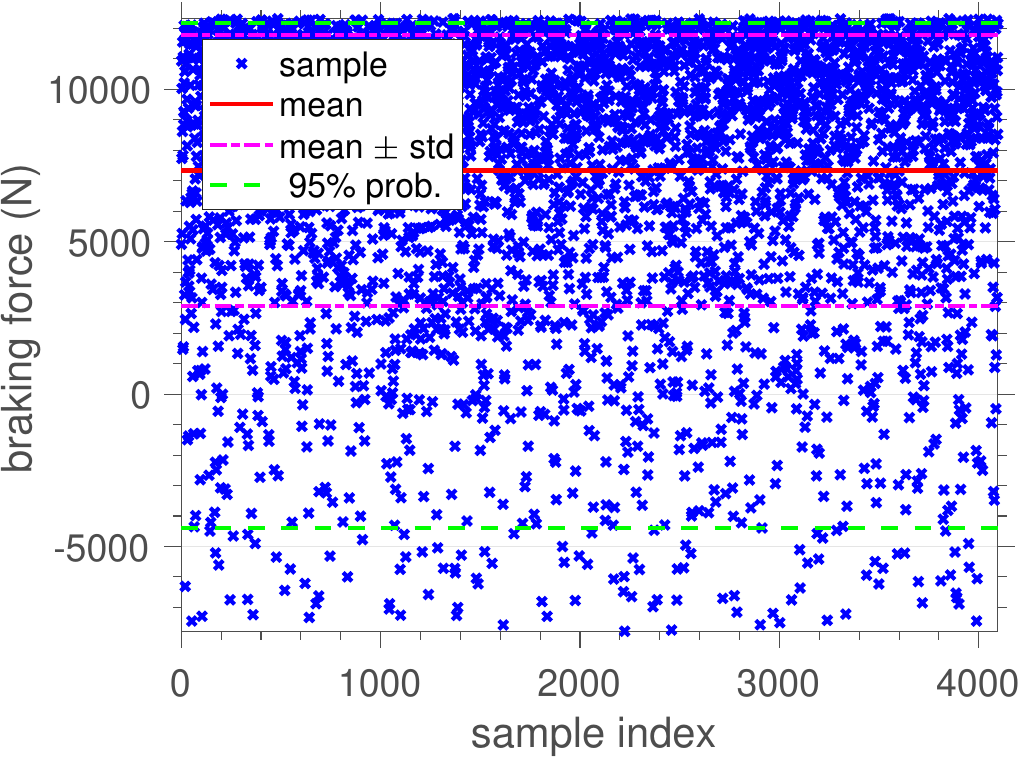} ~ 
\includegraphics[scale=0.5]{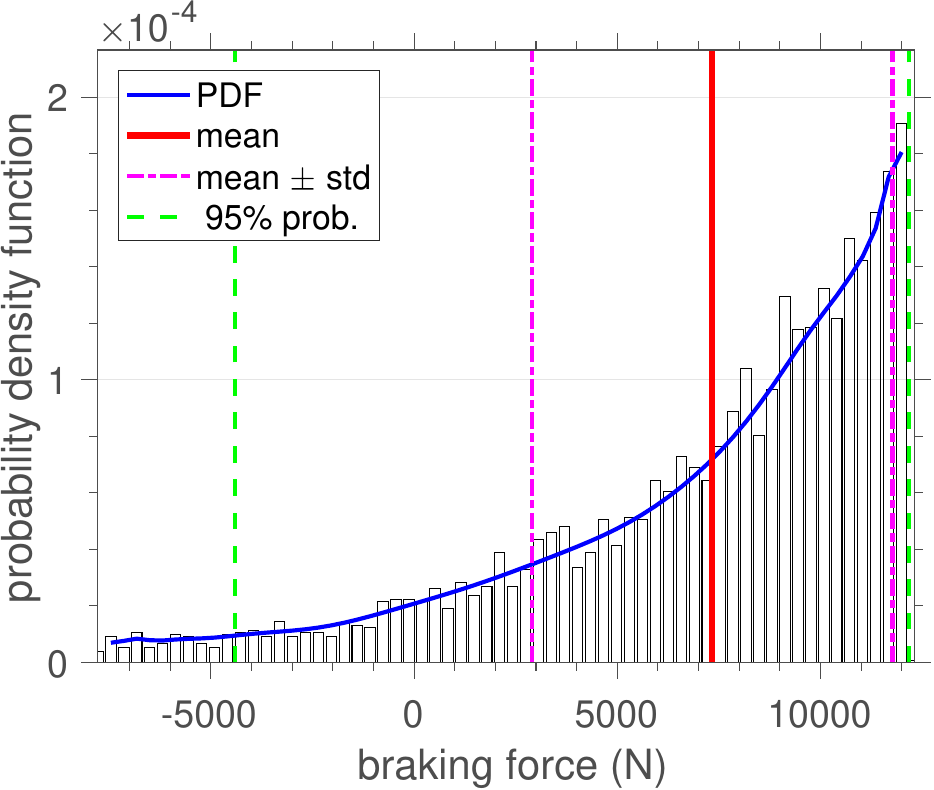}
\caption{Generated samples (top) and probability density function (bottom),
with statistical measures, for the braking force.}
\label{output_pdf_fig}
\end{figure}

\begin{figure}
\centering
\includegraphics[scale=0.5]{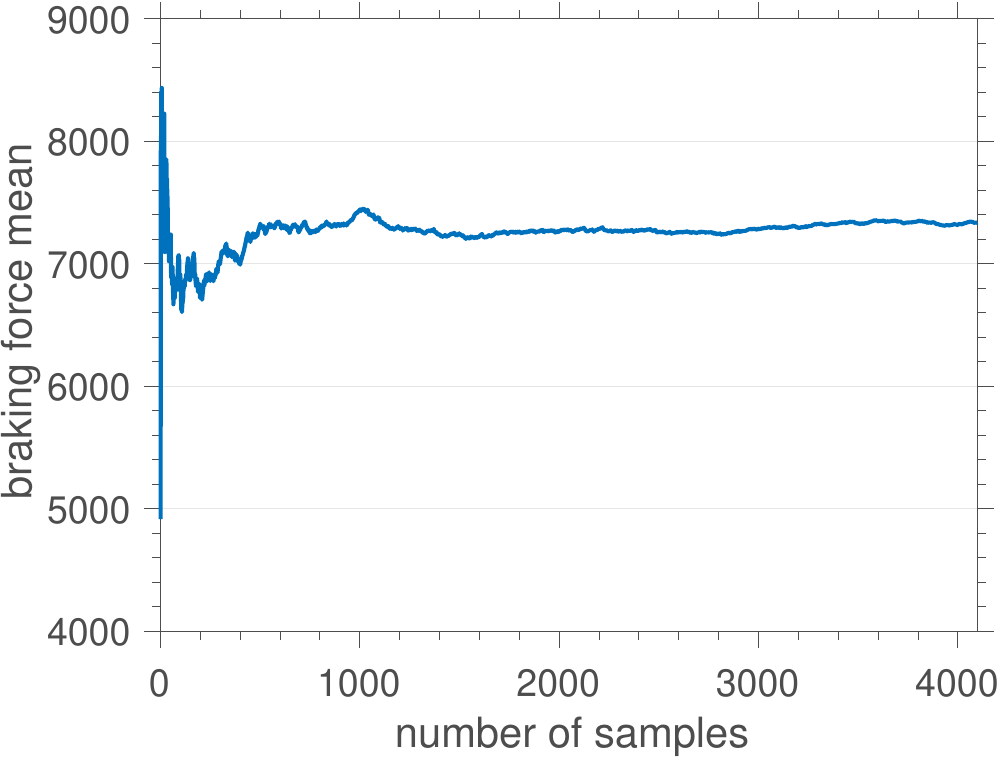} ~ 
\includegraphics[scale=0.5]{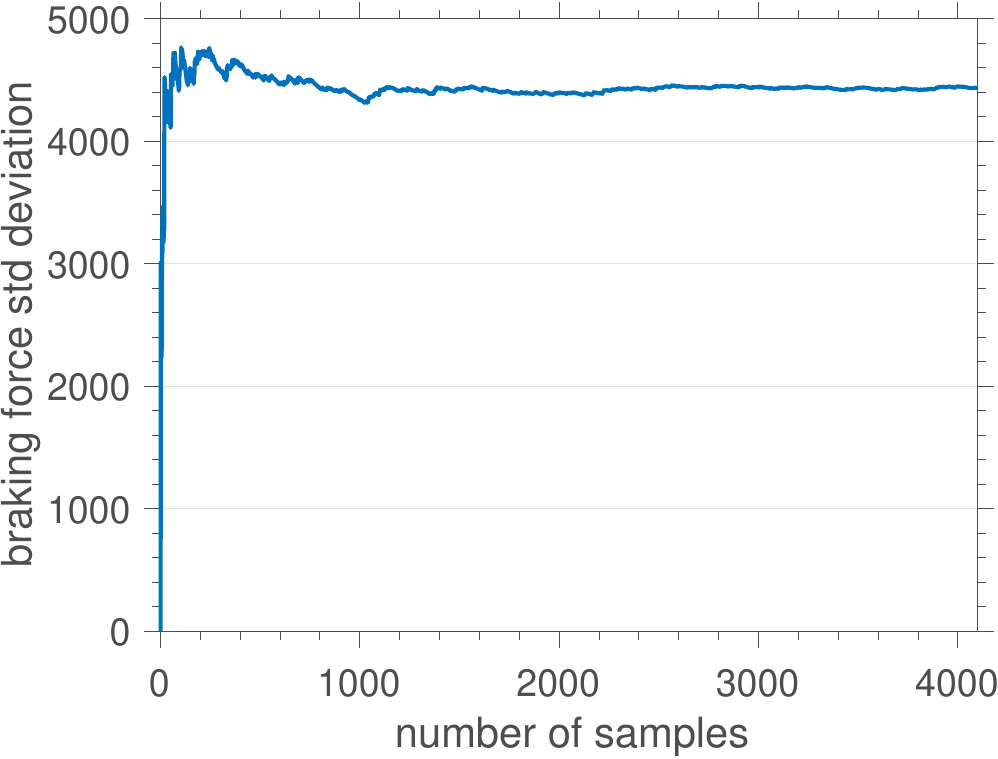}
\caption{Statistical estimators convergence for the braking force mean 
(top) and standard deviation (bottom).}
\label{output_conv_fig}
\end{figure}

The next step involves the model evaluation in each pair $(X_1, X_2)$ 
previously generated, which gives rise to the set of possible values 
for the braking force $F_h$, shown in {the} top part of Figure~\ref{output_pdf_fig}.
{In} the bottom part of the same figure the reader can observe a histogram that estimates 
the $F_h$ PDF form, as well as a nonparametric fitting obtained by a smooth 
curve. Mean, standard deviation, and a 95\% confidence interval can be
seen in both, top and bottom figures. To prove that these estimates are reliable, 
the authors also show the convergence of the mean and standard deviation estimators, 
as a function of the number of samples, in Figure~\ref{output_conv_fig}.

It may be noted that the PDFs of $F_h$ and $F_s$ have a very similar shape, 
suggesting that the mechanical model preserves the shape of the spring reaction 
force distribution. This result is at the least curious and unexpected, since the 
angular {dependencies} introduced in the mechanical model by Eqs.(\ref{N3_eq}) and (\ref{N1_eq}) 
define a structure of multiplicative uncertainty {between} $\alpha$ and $F_s$, what
should make $F_h$ not invariant with respect to the input distribution.

\begin{figure}
\centering
\includegraphics[scale=0.5]{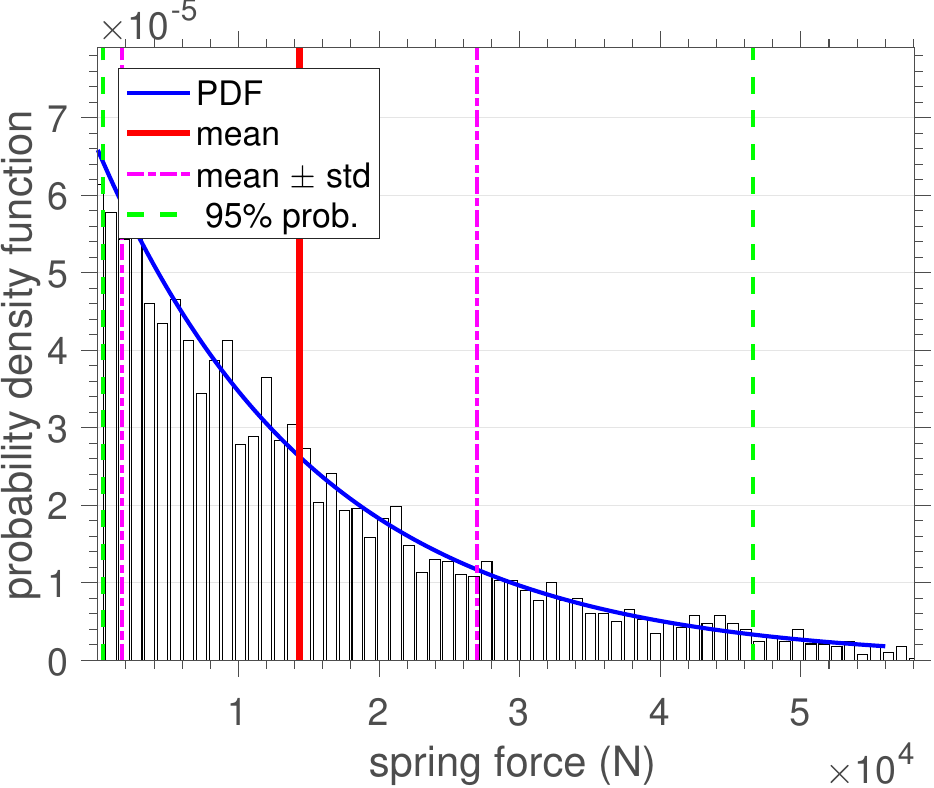} ~ 
\includegraphics[scale=0.5]{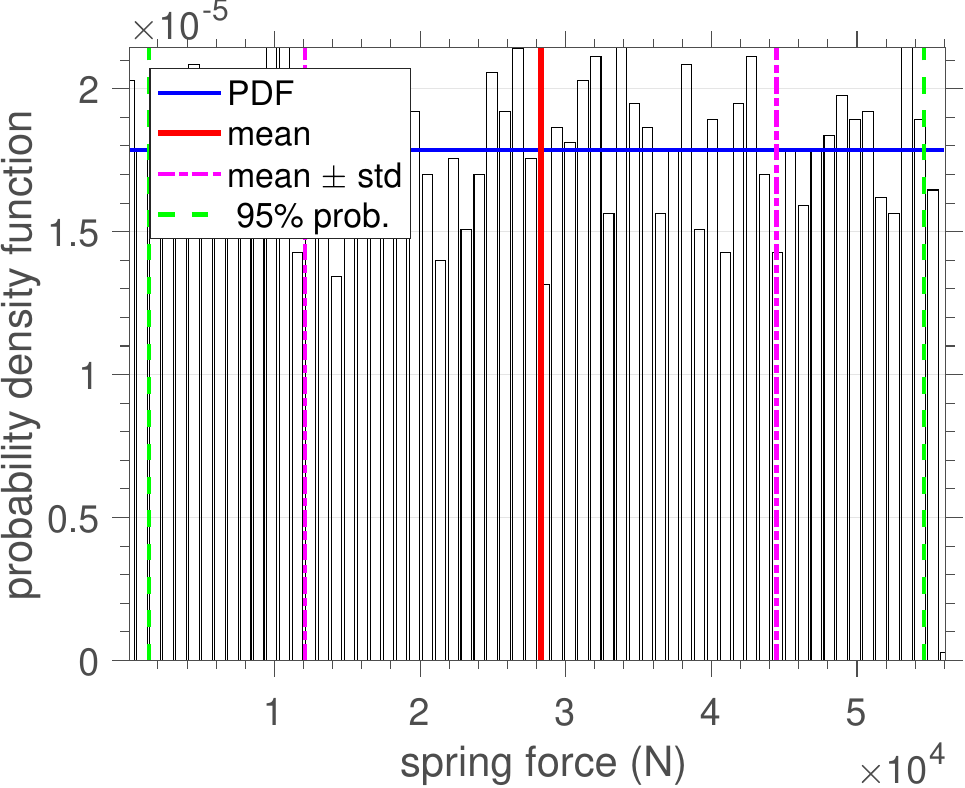} ~
\includegraphics[scale=0.5]{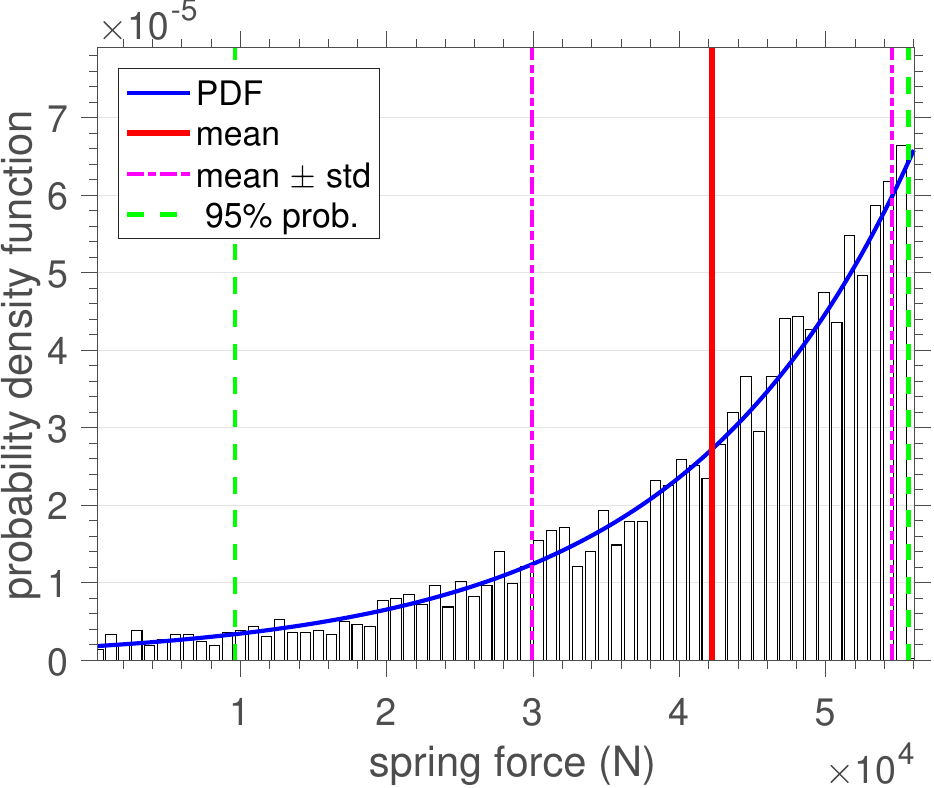}
\caption{Probability density functions and statistical measures 
for the spring force reaction with different mean values:
14 kN (top); 28 kN (middle); 42 kN (bottom).}
\label{Fs_pdfs_fig}
\end{figure}

\begin{figure}
\centering
\includegraphics[scale=0.5]{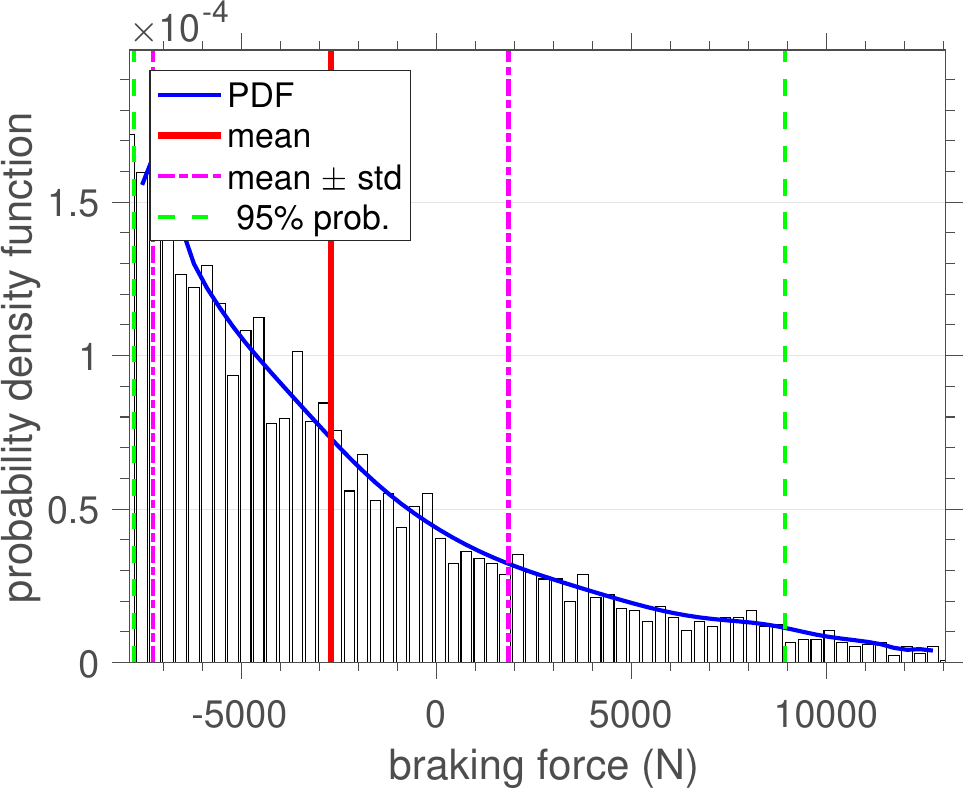} ~ 
\includegraphics[scale=0.5]{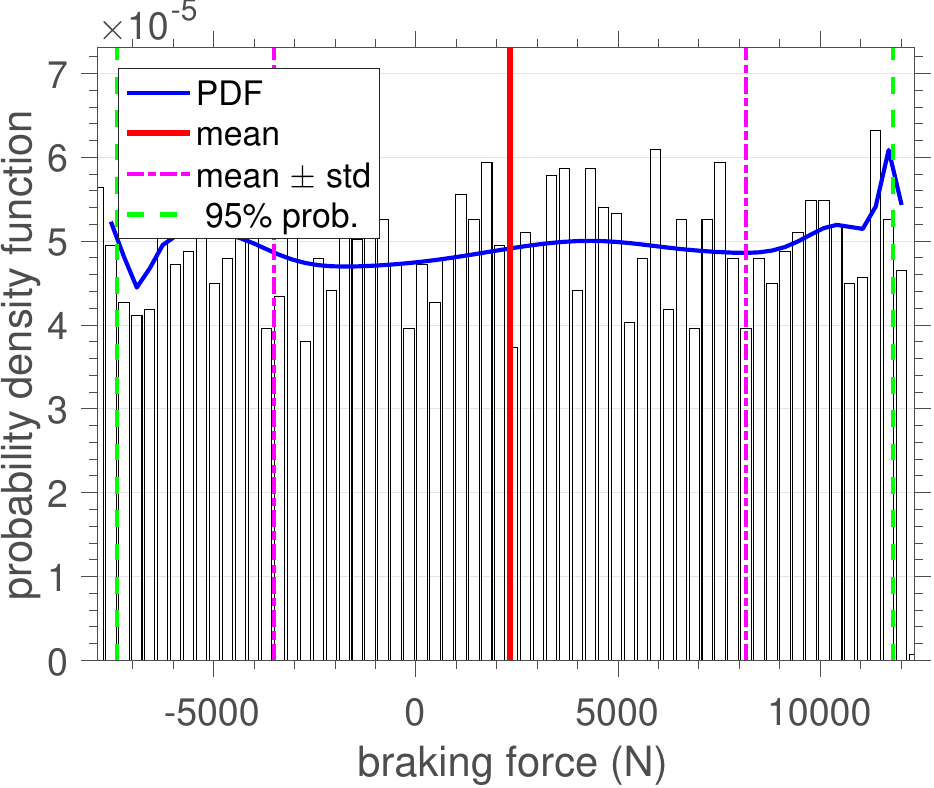} ~
\includegraphics[scale=0.5]{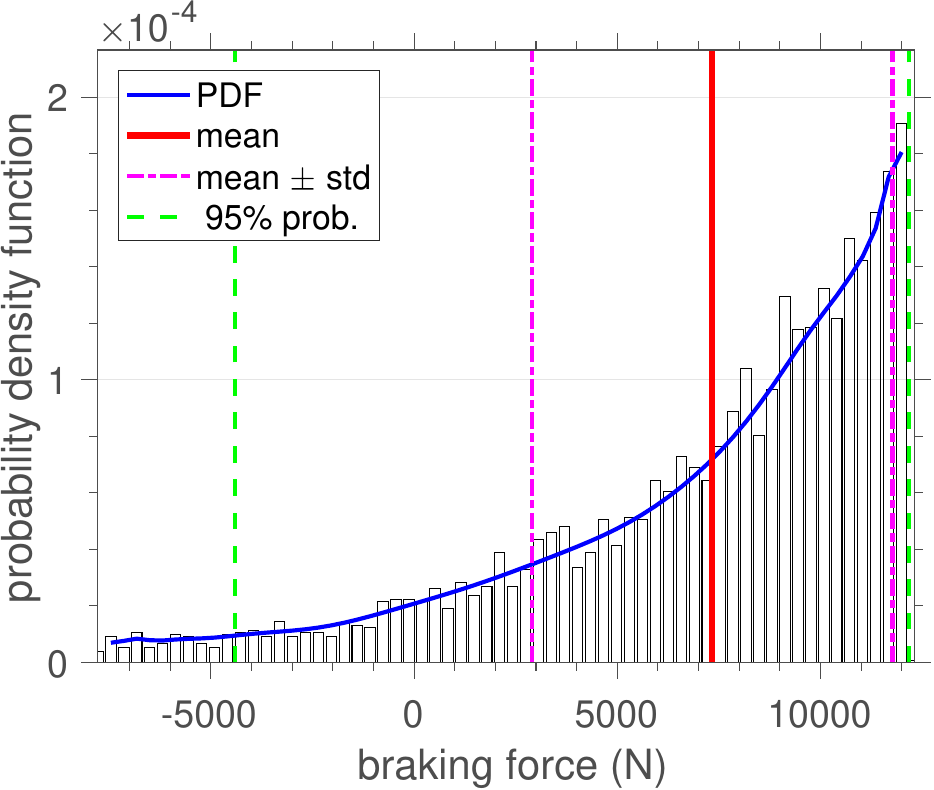}
\caption{Probability density functions and statistical measures 
for the braking force with different input mean values for $F_s$:
14 kN (top); 28 kN (middle); 42 kN (bottom).}
\label{Fh_pdfs_fig}
\end{figure}

This result suggests that the nonlinearity associated with the alpha parameter 
is very weak, which causes $F_h$ to behave as an affine map of $F_s$, 
and thus to preserve the form of its distribution. This hypothesis is reinforced 
by analyzing the system response by keeping $\alpha$ distribution and 
$F_s$ support fixed, while the mean value of the latter parameter is varied,
assuming the values equal to for 14 kN, 28 kN\footnote{For this value, 
which corresponds to the midpoint of the support, the distribution 
degenerates into an uniform.} and 42 kN. The probability densities 
corresponding to these different inputs, and the corresponding outputs 
of the mechanical system can be seen in 
Figures~\ref{Fs_pdfs_fig} and \ref{Fh_pdfs_fig}, respectively.
In all cases the input and output PDFs have the same shape.

The results of this study allow one  to conclude that uncertainties 
in $\alpha$ parameter does not have {significant} influence {on}
the braking force behavior. Simulations propagating only 
$\alpha$ uncertainties, not included here because of 
space limitation, demonstrate such an assertion.
However, the uncertainty propagation study 
also shows that the variability of $F_s$ {cannot} be 
ignored, since it has great influence on the statistical 
behavior of $F_h$.

\subsection{Optimization}

In this section the problem of optimum design of the brake system 
is addressed. The geometric dimensions $\bm{s} = (a, c)$ are used 
as design variables, considering as admissible region
$50 \leq a \leq 60$ mm and $50 \leq c \leq 55$ mm.

\begin{figure}
\centering
\includegraphics[scale=0.45]{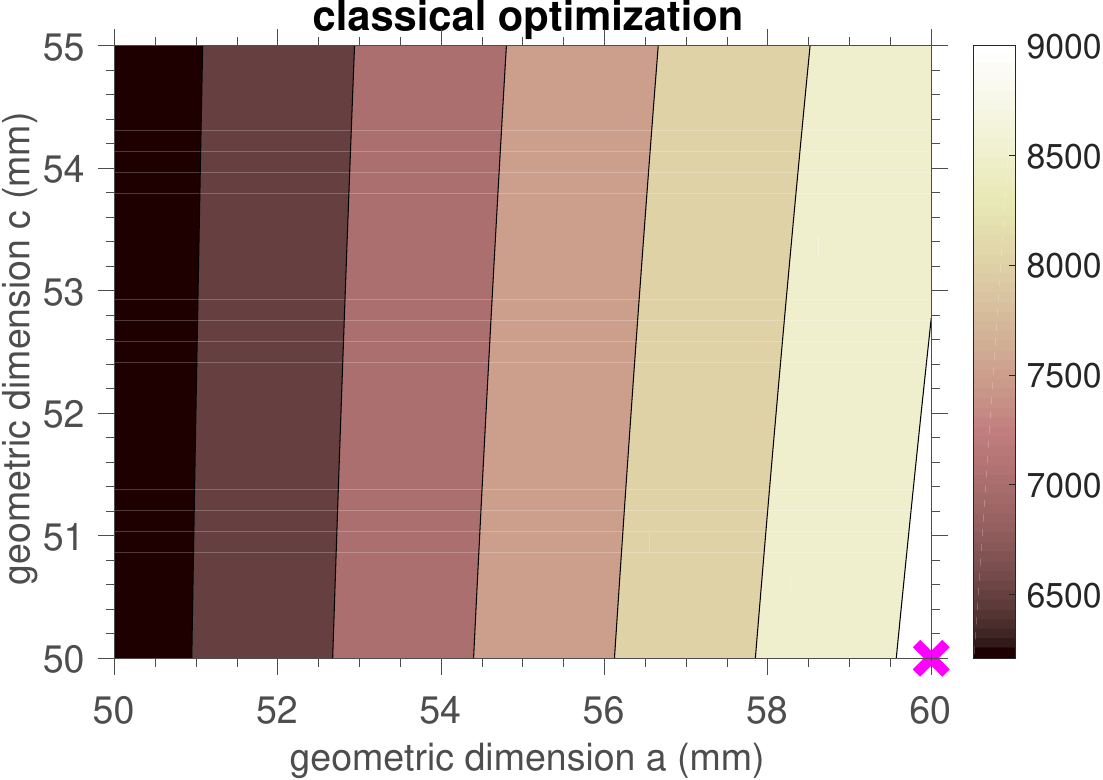} 
\caption{Contour map for the classical objective function and
with the optimum point is indicated by a cross.}
\label{classical_opt_fig}
\end{figure}

The optimization problem is solved using the standard 
sequential programming quadratic (SQP) algorithm
{obtained from Matlab (see chap. 18 of \cite{nocedal2006})},
being the contour map of the objective function (\ref{opt_eq1})
illustrated in Figure~\ref{classical_opt_fig}, 
which also highlights the optimum point found.

Despite {the fact that} this result {offers} a starting point for an optimal project for the brake system,
it does not take into account the effect of the uncertainties underlying the operating 
conditions, which can considerably affect the system response, as shown in the
previous section. In this way, robust optimization presents itself as a natural alternative.

For the robust optimization problem the design variables $\bm{s}=(a,c)$ are considered
once more, with the same ranges of admissible values used above. The uncertainties in $\alpha$ and 
$F_s$ are modeled as in section~\ref{stoch_model}, and the probabilistic constraint is 
characterized by the parameters $y^{\ast} = 0.5$ kN and $P_r = 5\%$. The convex weights 
$\beta_1 = \beta_2 = \beta_3 = 2/10$ and $\beta_4 = 4/10$ are adopted in the 
robust objective function.

This second problem is much more complex because the constraint to be satisfied 
is nonconvex, offering additional challenges to the numerical solution procedure.
But for the values described above the SPQ algorithm is able to find a solution.

\begin{figure}
\centering
\includegraphics[scale=0.45]{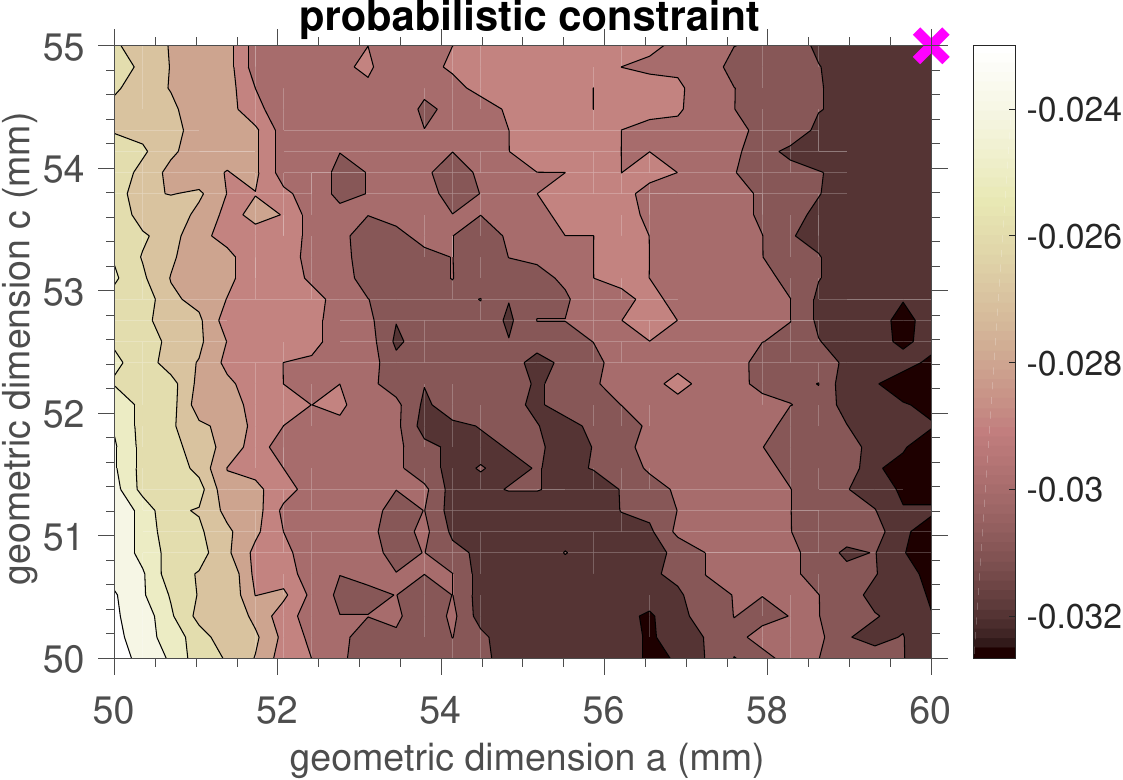} 
\caption{Contour map for probabilistic constraint with
the optimum point is indicated by a cross.}
\label{robust_opt_fig1}
\end{figure}

\begin{figure}
\centering
\includegraphics[scale=0.45]{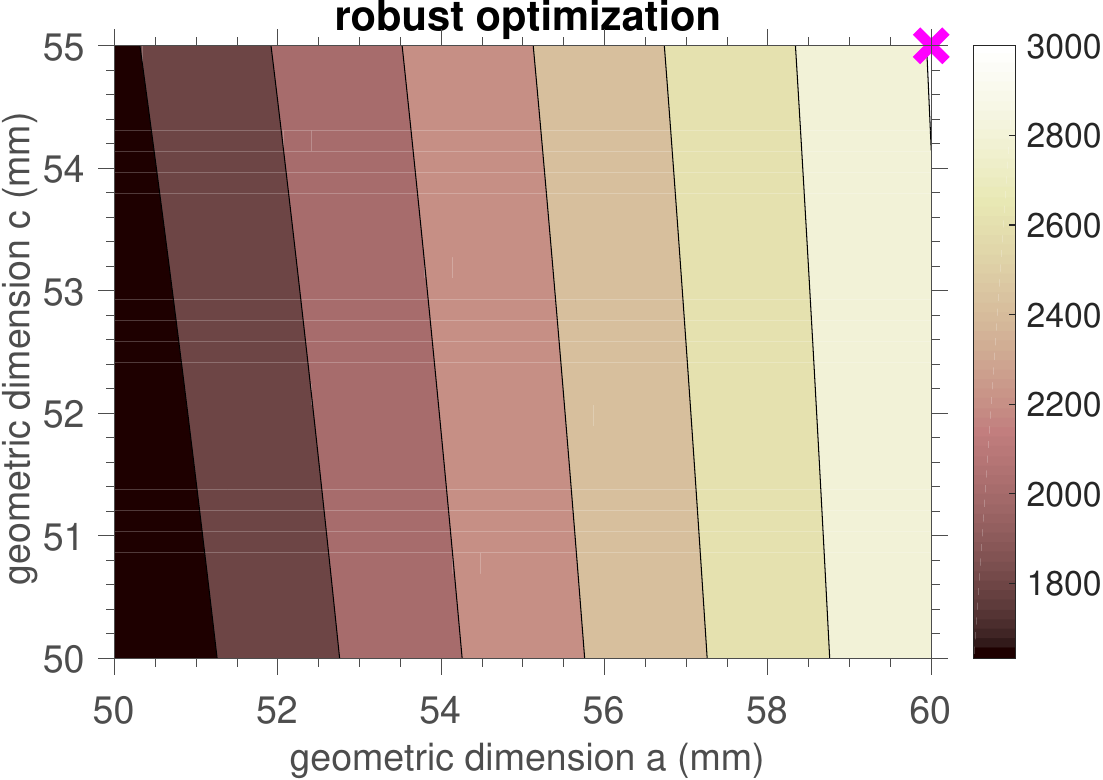} 
\caption{Contour map for the robust objective function with
the optimum point is indicated by a cross.}
\label{robust_opt_fig2}
\end{figure}

The reader can see the contour map of the probabilistic constraint (\ref{ropt_eq2})
in Figure~\ref{robust_opt_fig1}, while Figure~\ref{robust_opt_fig2} presents 
the contour map for the robust objective function (\ref{ropt_eq1}). Although 
the objective function still maintains a smooth appearance, the problem gains 
a non-convex status by the irregular forms of constraint.

Once the robust objective function takes into account other design criteria
than in (\ref{opt_eq1}), its behavior is different from the classical objective 
function shown in Figure~\ref{classical_opt_fig}, thus having a different 
optimal point. 

It is also worth noting that this second formulation of the optimization problem 
considers the effects of uncertainties, which in a realistic system are always 
present, thus offering a design option more suitable for projects that {cannot} 
ignore such variabilities.


\section{Summary and conclusions}
\label{concl_remaks}

This work presents a study regarding
the optimization and uncertainty quantification
of an elevator brake system. 
The paper starts from an original construction of 
a safety gear for the brake, for which a 
mechanical-mathematical model is
constructed. Studies involving the quantification of the 
braking force uncertainties due to the variability 
in the brake cam angle and the spring reaction force
are presented, showing that spring force uncertainties 
 are more significant. 
The {paper} also focuses on the optimal design of an elevator
brake system, showing through the solution of a robust 
optimization problem that operating conditions uncertainties 
can significantly influence its efficiency.


\section*{Funding}
{
This research was financed in the framework of the 
project Lublin University of Technology-Regional 
Excellence Initiative, funded by the Polish Ministry 
of Science and Higher Education (contract no. 030/RID/2018/19).
The fourth author acknowledge the financial support 
given by the Brazilian agencies 
Carlos Chagas Filho Research Foundation of 
Rio de Janeiro State (FAPERJ)
under grants E-26/010.002.178/2015 and 
E-26/010.000.805/2018 and
Coordena\c{c}\~{a}o de Aperfei\c{c}oamento de Pessoal 
de N\'{\i}vel Superior - Brasil (CAPES) - 
Finance Code 001.}

\section*{Compliance with ethical standards}

\section*{Conflict of Interest }
The authors declare that they have no conflict of interest.


\bibliographystyle{unsrtnat}


\end{document}